\documentclass[aps,prb,reprint,amsmath,amssymb,floatfix,citeautoscript]{revtex4-2}
\usepackage{graphicx}
\usepackage{color}
\usepackage{bm}
\usepackage{tabularx,booktabs,multirow,siunitx}

\usepackage{xr-hyper}
\usepackage[colorlinks,linkcolor=blue,citecolor=blue,urlcolor=blue]{hyperref}
\usepackage[noabbrev]{cleveref}

\creflabelformat{equation}{(#2#1#3)}
\crefname{figure}{Fig.}{Figs.}
\crefname{section}{Section}{Sections}

\usepackage{filecontents}

\newcommand*{\x}{\mskip-2mu\times\mskip-2mu}

\begin{document}

\title{Machine Learning Phonon Spectra for\\Fast and Accurate Optical Lineshapes of Defects}

\author{Mark E. Turiansky}
\email{mark.e.turiansky.ctr@us.navy.mil}
\affiliation{US Naval Research Laboratory, 4555 Overlook Avenue SW, Washington, DC 20375, USA}

\author{John L. Lyons}
\affiliation{US Naval Research Laboratory, 4555 Overlook Avenue SW, Washington, DC 20375, USA}

\author{Noam Bernstein}
\affiliation{US Naval Research Laboratory, 4555 Overlook Avenue SW, Washington, DC 20375, USA}

\date{\today}

\begin{abstract}
    The optical properties of defects in solids produce rich physics, from gemstone coloration to single-photon emission for quantum networks.
    Essential to describing optical transitions is electron-phonon coupling, which can be predicted from first principles but requires computationally expensive evaluation of all phonon modes in simulation cells containing hundreds of atoms.
    We demonstrate that this bottleneck can be overcome using machine learning interatomic potentials with negligible accuracy loss.
    A key finding is that atomic relaxation data from routine first-principles calculations suffice as a dataset for fine-tuning, though additional data can further improve models.
    The efficiency of this approach enables studies of defect vibrational properties with high-level theory.
    We fine-tune to hybrid functional calculations to obtain highly accurate spectra, comparing with explicit calculations and experiments for various defects.
    Notably, we resolve fine details of local vibrational mode coupling in the luminescence spectrum of the T center in Si, a prominent quantum defect.
\end{abstract}

\maketitle

The interaction of photons with native defects or impurities in crystalline materials is the basis of various phenomena and technologies.
Absorption of light by impurities is responsible for the coloration of gemstones~\cite{fritsch_update_1987}.
The luminescence of Cr impurities in sapphire (Al$_2$O$_3$), known as ruby, enabled the first solid-state lasers~\cite{maiman_stimulated_1960}.
Currently, defects are being explored as the basis of quantum technologies~\cite{dreyer_first-principles_2018,wolfowicz_quantum_2021};
for example, isolated defects may act as single-photon emitters~\cite{turiansky_rational_2024}, enabling quantum networks~\cite{northup_quantum_2014}.

A native defect or impurity is not isolated from the crystalline lattice but interacts with the bath of phonon modes.
(We will refer to native defects and impurities collectively as defects.)
Coupling between the electronic states of the defect and the phonons of the lattice manifests as the phonon sideband observed during an optical transition~\cite{stoneham_theory_1975}.
First-principles calculations based on density functional theory (DFT) have been an invaluable tool for the study of defects and their coupling to phonons~\cite{freysoldt_first-principles_2014,dreyer_first-principles_2018}.
The formulation to quantitatively evaluate luminescence lineshapes from first principles was pioneered by Alkauskas \textit{et al}~\cite{alkauskas_first-principles_2014-1}.
Such calculations are computationally demanding, requiring thousands of DFT calculations to evaluate phonons of supercells containing hundreds of atoms.
Often, this prevents the utilization of the computationally demanding exchange-correlation functionals that are necessary to quantitatively study defects.

Machine learning interatomic potentials (MLIPs) have emerged as a promising route to accelerate first-principles calculations~\cite{wang_scientific_2023,chen_universal_2022,batatia_foundation_2024}.
In short, MLIPs describe the total energy as a function of atomic coordinates by utilizing highly flexible machine-learning techniques (e.g., neural networks) to achieve the mapping.
Application of MLIPs typically requires data-intensive training for the system of interest.
An exciting development is the surprising generalizability of MLIPs when pre-trained on large, diverse datasets (referred to as foundation models or universal MLIPs)~\cite{chen_universal_2022,batatia_foundation_2024,loew_universal_2025}.
Relatively small datasets can be used to refine these models when increased accuracy is desired~\cite{deng_systematic_2025,kaur_data-efficient_2025}, in particular for vibrational properties~\cite{pota_thermal_2024,loew_universal_2025}.

While MLIPs show promise for defect studies---including accelerated atomic relaxation~\cite{mosquera-lois_machine-learning_2024}, finite-temperature effects~\cite{mosquera-lois_point_2025}, and high-throughput screening~\cite{kavanagh_identifying_2025}---optical property prediction remains challenging.
Linder\"{a}lv \textit{et al.} demonstrated the first application of a MLIP to optical lineshapes~\cite{linderalv_optical_2024}.
However, the use of classical autocorrelation functions limits their approach to relatively high temperatures, and intensive system-specific training restricted their functional choice to a semi-local functional.
Sharma \textit{et al.}~\cite{sharma_accelerating_2025} investigated the use of foundation models within the formalism of Alkauskas \textit{et al.} but compared with PBE calculations, and even so, they found large discrepancies for strong coupling and stopped short of any fine-tuning.
These limitations hinder quantitative analysis and experimental comparisons.

Indeed, an underlying limitation of foundation models is that they are typically trained on semi-local DFT calculations.
For defects, this is a particularly severe constraint.
The inaccuracy of semi-local functionals is tied to a failure to accurately describe charge localization~\cite{cohen_insights_2008}.
Defective systems require a simultaneous description of localized defect states and delocalized bulk states.
In the extreme case, semi-local functionals may miss the existence of deep defect states, particularly for polaronic distortions~\cite{wickramaratne_assessing_2024}.

Here, we demonstrate MLIPs that are capable of producing highly accurate optical spectra, which are quantitatively compared with experiments.
We first show that a foundation model can produce qualitatively correct optical spectra when used in conjunction with the atomic relaxation determined by hybrid functional DFT.
Carrying out the atomic relaxation of a defect produces a small dataset, which we find to be sufficient for fine-tuning of the MLIP.
This enables a significant improvement in the accuracy of the produced spectra essentially for free:
no additional DFT calculations are needed, and training on a small dataset requires less than an hour of calculation time on a GPU.
We investigate ways to generate additional data to further improve the model, demonstrating that as few as ten additional calculations can further improve performance.
To show the effectiveness of our approach, we apply the method to predict the optical spectra of several defects actively being studied in the literature.

\section*{Results}
To demonstrate our approach, we will focus on evaluating the normalized luminescence spectrum $L(\hbar\omega)$ (Methods).
Luminescence is one of the many experimental techniques that can be used to characterize a given defect and for which theoretical techniques exist.
Our approach could also be applied to explore absorption spectra or could be easily extended to account for Jahn-Teller effects~\cite{razinkovas_vibrational_2021}.

We benchmark against explicit DFT calculations in three cases (Methods): a substitutional carbon impurity (C$_{\rm N}$) in GaN~\cite{lyons_carbon_2010}, the nitrogen-vacancy (NV) center in diamond~\cite{alkauskas_first-principles_2014-1}, and a carbon dimer (C$_{\rm B}$-C$_{\rm N}$) in hBN~\cite{mackoit-sinkeviciene_carbon_2019}.
These cases are technologically relevant, span a range of electron-phonon coupling strengths, and include polar and non-polar materials, as well as a two-dimensional material.
For the NV center and carbon dimer, we study the lowest energy spin-conserving internal transition.
In contrast, we study the capture of a hole from the valence band at C$_{\rm N}$ in GaN, during which the defect changes from the negative to neutral charge state.

\subsection*{Foundation Model}
\begin{figure*}[htb!]
    \centering
    \includegraphics[width=\textwidth,height=\textheight,keepaspectratio]{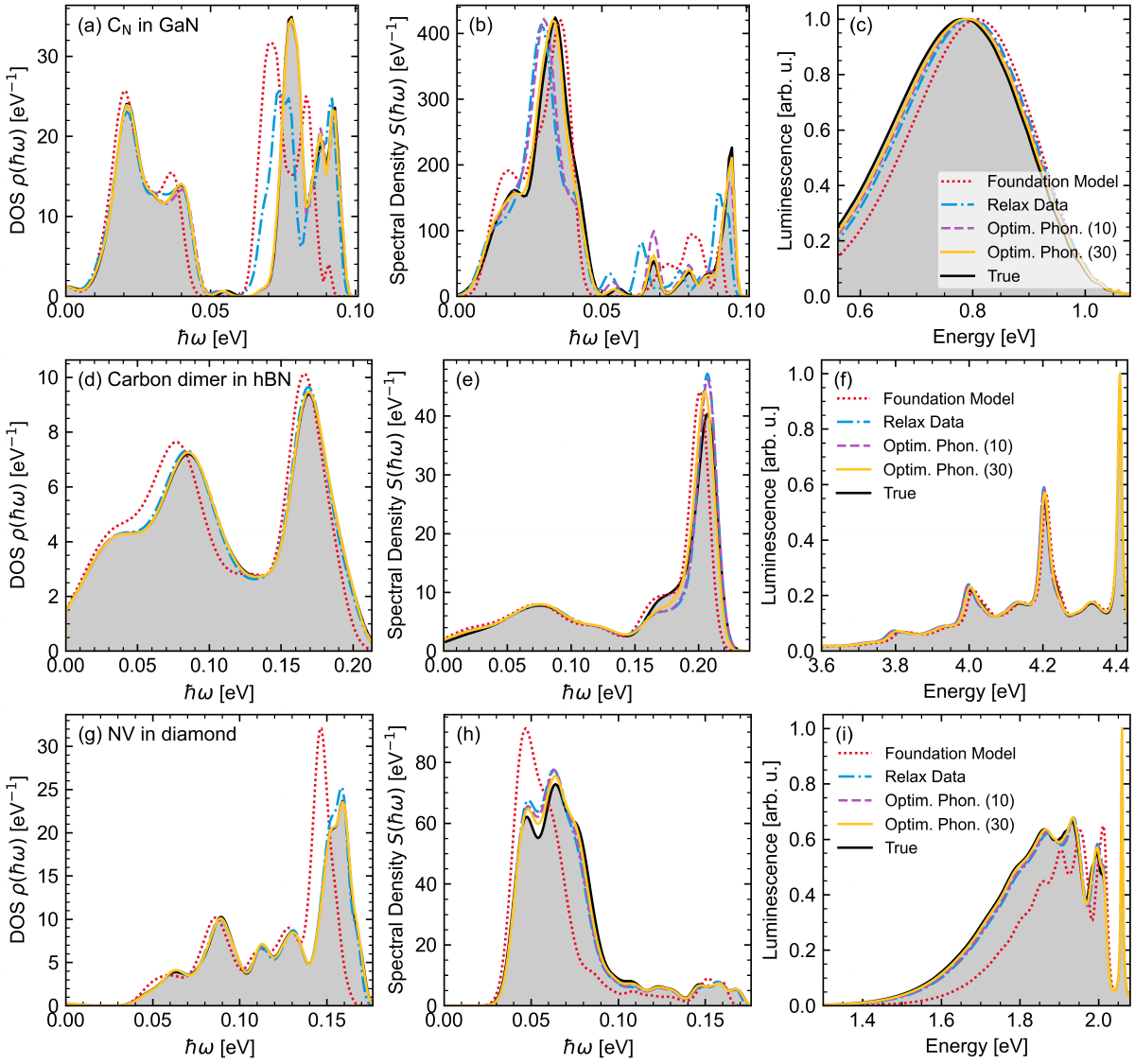}
    \caption{\label{fig:bench_pl}
        The phonon density of states (left), spectral density (middle), and luminescence spectrum (right) for C$_{\rm N}$ in GaN (first row), the carbon dimer in hBN (second row), and the NV center in diamond (third row).
        The black, shaded region corresponds to the results explicitly calculated with DFT.
        The foundation model is shown in red (dotted), while the fine-tuned model using relaxation data is shown in blue (dash, dotted).
        Fine-tuning with $N_{\rm conf} = 10$ (purple, dashed) and 30 (yellow, solid) additional configurations generated with the optimized phonon method are also shown.
    }
\end{figure*}

We employ the atomic cluster expansion with message passing (MACE)~\cite{batatia_mace_2022,batatia_design_2025} MLIP (Methods).
First, we assess the \texttt{mace-omat-0-medium} foundation model~\cite{batatia_foundation_2024} for approximating each defect's vibrational properties.
In \cref{fig:bench_pl}, we show the phonon density of states $\rho(\hbar\omega)$, spectral density $S(\hbar\omega)$, and luminescence $L(\hbar\omega)$ for the three benchmark cases (Methods).
As the foundation model is trained on semi-local DFT, we should not expect it to exactly reproduce the vibrational spectrum of hybrid functional DFT.
This is clearly reflected in the phonon density of states [\cref{fig:bench_pl}(a,d,g)] on which $S(\hbar\omega)$ [\cref{fig:bench_pl}(b,e,h)] and $L(\hbar\omega)$ [\cref{fig:bench_pl}(c,f,i)] implicitly depend.
While the qualitative structure of $\rho(\hbar\omega)$ and $S(\hbar\omega)$ is captured in all cases, there are differences, particularly for high frequency modes.
A technical analysis of the spectra is given in Supplementary Information \cref{sec:fm_bench}.
We find that the foundation model---when used in conjunction with the atomic relaxation vector from HSE calculations---is capable of qualitatively correct predictions of the luminescence, but for many applications, further refinement would be desirable.

\subsection*{Atomic Relaxation Dataset}
When studying a defect, an atomic relaxation is performed to identify the equilibrium geometry.
This atomic relaxation produces a small dataset of configurations that can be used for fine-tuning (Methods).
The resulting spectra from fine-tuning to the relaxation dataset are shown in \cref{fig:bench_pl} (labeled as ``Relax Data'').
A significant improvement in the predicted spectra is obtained, achieving accuracy close to that of the benchmark DFT calculations.
This observation adds to the mounting evidence that fine-tuning of foundation models can be performed with surprisingly little data~\cite{deng_systematic_2025,kaur_data-efficient_2025,pota_thermal_2024}.

This approach offers a significant computational advantage thanks to the effectively free dataset:
an atomic relaxation is always performed, and no additional DFT calculations were needed.
On an NVIDIA A100 GPU, training takes less than an hour, and analytical evaluation of the dynamical matrix takes approximately two minutes.
These numbers can be contrasted with an explicit hybrid DFT calculation of the dynamical matrix, which requires evaluating 576 configurations for C$_{\rm N}$ in GaN, 1,296 for the NV center in diamond, and 1,440 for the carbon dimer in hBN, with each configuration requiring approximately 15 minutes on the GPU.
Thus, orders of magnitude of computational effort are saved.

Ultimately, these results are satisfying, as they enable a quantitative comparison with experiments, but some studies demand a more exact reproduction of the true DFT calculation.
We investigate techniques to augment the atomic relaxation dataset in the following.

\subsection*{Additional Data}
Several approaches for generating additional configurations are assessed and compared in Supplementary Information \cref{sec:ad_bench}.
We generate 10 and 30 additional configurations for each defect (Methods).
In \cref{fig:bench_pl}, we show the resulting spectra for the three benchmark cases (labeled as ``Optim. Phon.'').
A significant improvement in the predicted quantities is obtained in all three cases, resolving many of the remaining discrepancies in the spectra.
The density of states, in particular, is extremely well captured.
In all of the benchmark cases, the predicted luminescence matches the one explicitly calculated with DFT.

Although including additional configurations requires extra DFT calculations, this still constitutes a major speedup.
Given that training the MACE model is a negligible expense, including 10 (30) modes constitutes a 57.6$\x$ (19.2$\x$) speedup in the case of C$_{\rm N}$ in GaN, 129.6$\x$ (43.2$\x$) speedup for the NV center in diamond, and 144$\x$ (48$\x$) speedup for the carbon dimer in hBN with an insignificant loss in prediction accuracy.

\subsection*{Application}
\label{sec:app}
We demonstrated the baseline performance of foundation models for the prediction of the vibrational properties of defects and formulated a systematic approach to further improve the model.
The approach achieves quantitative accuracy when benchmarked against explicit hybrid DFT calculations, and based on this, we argue that this approach enables quantitative comparisons with experiments that were not previously feasible.
To show this explicitly, we now study four ``blind'' applications of our approach to defects in complex materials (where blind refers to the absence of any DFT benchmarking data).

\subsubsection*{N$_{\rm O}$ in ZnO}
ZnO is a wide-bandgap oxide utilized in electronic and optoelectronic devices~\cite{janotti_fundamentals_2009}.
Defects in ZnO have been the subject of research interest, in particular for doping.
N impurities were once thought to potentially lead to $p$-type doping in ZnO, and first-principles calculations were essential in correcting this notion~\cite{lyons_why_2009}.
Subsequent work suggested N$_{\rm O}$ to be the origin of luminescence near 1.7~eV---a consequence of its deep acceptor nature~\cite{tarun_nitrogen_2011}.
However, a first-principles evaluation of the luminescence spectrum was never attempted.

\begin{figure}[htb!]
    \centering
    \includegraphics[width=\columnwidth,height=\textheight,keepaspectratio]{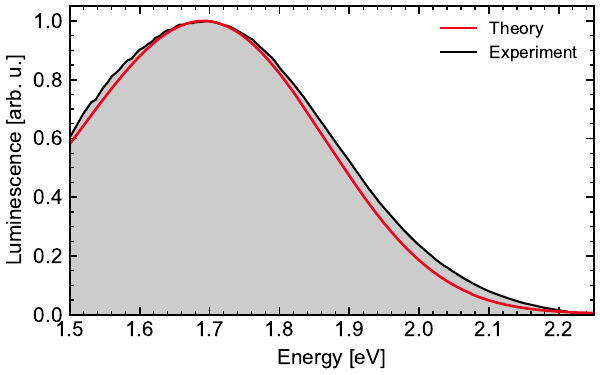}
    \caption{\label{fig:ZnO}
        Luminescence spectrum of N$_{\rm O}$ in ZnO at $T = 300$~K.
        The red line is the calculated spectrum, and the black line is the experimental spectrum extracted from Ref.~\onlinecite{tarun_nitrogen_2011}.
        We shift the calculated spectrum to account for typical DFT errors, aligning the peak intensity, and normalize both spectra to the peak intensity.
    }
\end{figure}

In \cref{fig:ZnO}, we use a model fine-tuned to relaxation data to compute the luminescence spectrum of N$_{\rm O}$ in ZnO, resulting from radiative capture of holes.
We find excellent agreement between the calculated and experimental luminescence spectra, further supporting the attribution to N$_{\rm O}$ impurities.
This center has relatively large coupling to phonons [$S_{\rm tot} = 12.6$, \cref{eq:Stot}], explaining the broad luminescence lineshape.

\subsubsection*{$V_{\rm Si}V_{\rm C}$ in 4H-SiC}
\begin{figure}[htb!]
    \centering
    \includegraphics[width=\columnwidth,height=\textheight,keepaspectratio]{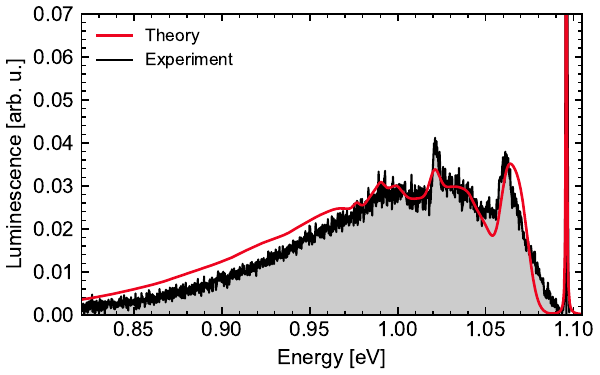}
    \caption{\label{fig:SiC}
        Luminescence spectrum of $V_{\rm Si}V_{\rm C}$ in SiC at $T = 10$~K.
        The red line is the calculated spectrum, and the black line is the experimental spectrum from Ref.~\onlinecite{jin_photoluminescence_2021}.
        We shift the calculated spectrum to align the ZPL.
        The experimental spectrum is normalized to approximately match the phonon sideband of the calculation.
    }
\end{figure}

SiC is a mature electronic material utilized for a variety of applications and has been gaining attention as a host for quantum defects~\cite{son_developing_2020}.
One such quantum defect is the divacancy center $V_{\rm Si}V_{\rm C}$ in the 4H polymorph of SiC~\cite{christle_isolated_2017}.
The luminescence of the divacancy has been studied extensively~\cite{jin_photoluminescence_2021} and thus serves as a useful test case.
(We study the $kk$ orientation of the divacancy.)
In \cref{fig:SiC}, we show the luminescence predicted using a model fine-tuned on the atomic relaxation data and find excellent agreement with the experimental spectrum.
As a case of moderate electron-phonon coupling ($S_{\rm tot} = 3.17$), there are more fine details present in the phonon sideband relative to a case of strong electron-phonon coupling, and these details are well resolved by our approach.

\subsubsection*{Bi$_{\rm Pb}$ in CsPbBr$_3$}
CsPbBr$_3$ is from the family of halide perovskites, which are being actively pursued for optoelectronic devices~\cite{dey_state_2021}, especially for photovoltaics where the efficiencies rival state-of-the-art materials.
As an all-inorganic halide perovskite, CsPbBr$_3$ exhibits better stability than the hybrid halide perovskites and is particularly attractive for x- and gamma-ray detection~\cite{pan_ultrahigh-flux_2023}.
Recent work has connected Bi impurities with near-infrared photoluminescence~\cite{brittman_near-infrared_2025}.

\begin{figure}[htb!]
    \centering
    \includegraphics[width=\columnwidth,height=\textheight,keepaspectratio]{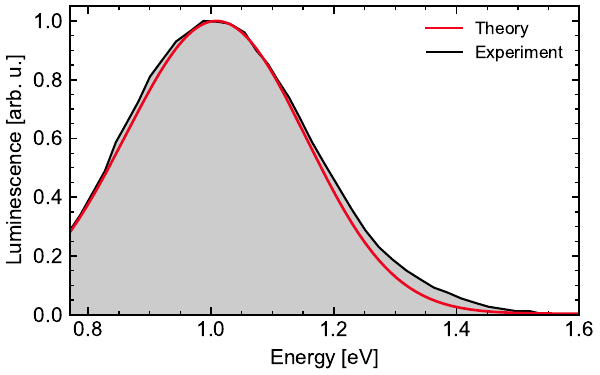}
    \caption{\label{fig:CsPbBr3}
        Luminescence spectrum of Bi$_{\rm Pb}$ in CsPbBr$_3$ at $T = 14$~K.
        The red line is the calculated spectrum, and the black line is the experimental spectrum extracted from Ref.~\onlinecite{brittman_near-infrared_2025}.
        We shift the calculated spectrum to align with the peak intensity and normalize both spectra to the peak intensity.
    }
\end{figure}

We calculate the luminescence spectrum of Bi$_{\rm Pb}$ in CsPbBr$_3$---again using a model fine-tuned to the relaxation data---shown in \cref{fig:CsPbBr3}, and find good agreement with the experimental spectrum.
This case provides an important test for our approach:
the elements involved explore regions of the periodic table that were not covered in the other tests.
In addition, Bi$_{\rm Pb}$ could be considered a case of ``extreme'' electron-phonon coupling.
We calculate $S_{\rm tot} = 84.1$, significantly larger than the other defects explored in this study.

In cases of such large electron-phonon coupling, anharmonicity may become a concern~\cite{zhang_correctly_2020,linderalv_optical_2024};
we suspect that some portion of the broadening in the luminescence may be a result of anharmonicity.
A rigorous treatment of anharmonicity and its effects on luminescence is beyond the scope of this study.
Still, the good agreement between the calculated and experimental spectra lends support to the attribution to Bi$_{\rm Pb}$ impurities.

\subsubsection*{T Center in Si}
The T center in Si is a promising quantum defect with a telecom-wavelength spin-photon interface~\cite{bergeron_silicon-integrated_2020} and is the subject of commercial efforts to realize quantum computing~\cite{afzal_distributed_2024}.
Si is an appealing host material for a variety of reasons, including the ability to obtain isotopically purified material with low defect densities.
While clearly advantageous for quantum applications, this high purity enables the measurement of luminescence spectra with unprecedented precision~\cite{karaiskaj_photoluminescence_2001}.
First-principles calculations have enhanced our understanding of the T center~\cite{dhaliah_first-principles_2022}, but a calculation of the luminescence matching the precision of experiments does not exist.
Towards this aim, we apply our approach to evaluate the lineshape function, uncovering the fine details of coupling to phonons.
Here we specifically address supercell scaling to compute the spectra in a 8000-atom supercell (Methods).
Broader comments on supercell scaling are found in the Discussion section.

\begin{figure*}[htb!]
    \centering
    \includegraphics[width=\textwidth,height=\textheight,keepaspectratio]{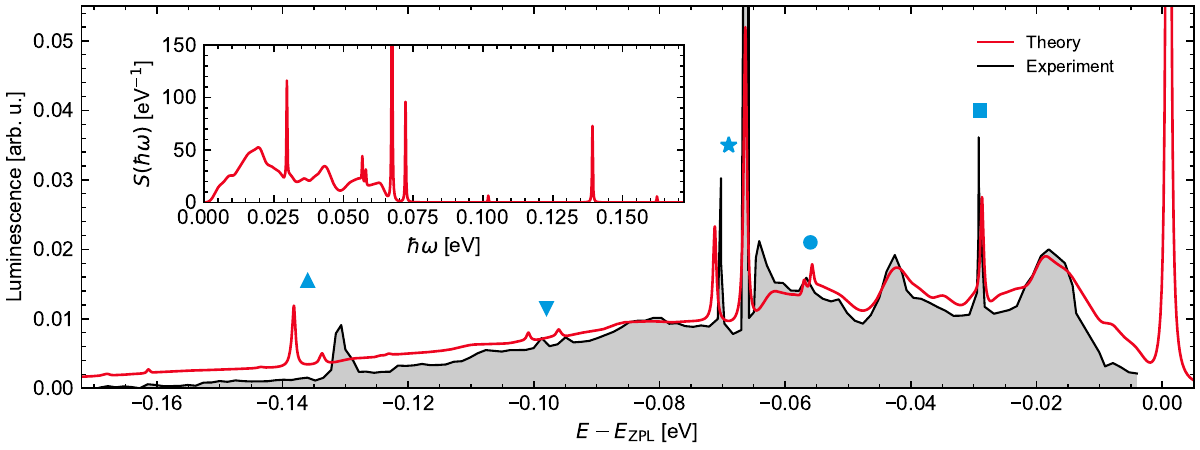}
    \caption{\label{fig:Tcenter}
        Luminescence spectrum of the T center in Si at $T = 4.2$~K.
        The red line is the calculated spectrum, and the black line is the experimental spectrum extracted from Ref.~\onlinecite{bergeron_silicon-integrated_2020}.
        We shift the calculated spectrum to align with the experimental ZPL.
        The intensity of the experimental spectrum is scaled to approximately match the intensity of the calculated phonon sideband.
        Features in the spectrum related to local vibrational modes are labeled with blue symbols.
        The calculated spectral density is shown in the inset.
    }
\end{figure*}

The calculated luminescence spectrum is shown in \cref{fig:Tcenter} and compared with the experimental spectrum of Ref.~\onlinecite{bergeron_silicon-integrated_2020}.
We find good agreement between the two: the important characteristics of the phonon sideband, including broader features and the sharp local vibrational modes of the defect, are well captured.
Two sharp peaks near $\approx$70~meV below the ZPL (labeled with a star) are one-phonon replicas of local vibrational modes.
These local vibrational modes involve the motion of less than 10 atoms of the defect and fall just outside the continuum of bulk modes (inset of \cref{fig:Tcenter}).
In contrast, the features at 29~meV and 56~meV below the ZPL (labeled with a square and circle, respectively) are one-phonon replicas of quasi-local vibrational modes.
These two modes fall within the continuum of bulk modes and correspond to the motion of $\approx$100-200 atoms.

Two pairings of doublet features are marked by triangles.
In both cases, the peaks further from the ZPL correspond to a one-phonon replica of a local vibrational mode where only one or two atoms are in motion.
The feature at 96~meV (down triangle) is a combination of the local vibrational mode at 29~meV (square) and at 67~meV (star), while the feature at 134~meV (up triangle) is a double of the 67~meV mode (star).
These observations were confirmed by selectively removing each mode from the calculation.

In total, our calculations have uncovered the interplay of bulk and localized vibrational modes in the T center's luminescence spectrum.
They demonstrate the power of our approach in accelerating our ability to build a first-principles understanding of defects.
We performed only 40 hybrid DFT calculations in a 512-atom supercell, in addition to the atomic relaxation.
Computing the dynamical matrix of the 512-atom defect supercell explicitly would require $2 \times 3 \times 512 = 3,072$ hybrid functional DFT calculations (a 76.8$\x$ speedup).
An explicit DFT evaluation of the dynamical matrix in the 8000-atom defect supercell is prohibitively expensive.
In contrast, approximately 5 hours of GPU time were required for our approach, primarily to compute the dynamical matrix of the 8000-atom cell.

\section*{Discussion}
\label{sec:discuss}
We have examined several different defect and host material combinations.
While not exhaustive, we can speculate on the generality of the approach across diverse materials classes.
The strength of electron-phonon coupling is proportional to the square-root mass of the host lattice (Supplementary Information \cref{sec:stot_prop}).
When electron-phonon coupling is strong, the luminescence is broad, and fine details are washed out.
One concludes that lineshape prediction of materials with heavy elements, deep in the periodic table, will be easier.
This is serendipitous, given that these elements tend to be less well represented in typical foundation-model training sets.

With the exception of the T center in Si, all DFT and MLIP calculations were performed in a fixed-size supercell.
This may seem surprising, given that a major advantage of MLIPs is the ability to address system sizes beyond what is tractable for DFT.
Even without discussing MLIPs, addressing the supercell scaling of defects is challenging.
The embedding scheme of Alkauskas \textit{et al.}~\cite{alkauskas_first-principles_2014-1} for large-supercell calculations of luminescence is only formally justified for non-polar materials.
While correcting the energy of a charged defect in periodic boundary conditions is standard~\cite{freysoldt_first-principles_2014,komsa_finite-size_2012}, correcting for the influence of charge on forces is relatively unexplored~\cite{razinkovas_vibrational_2021}.

In addition, most current MLIPs such as MACE have no notion of charge and are inherently short-ranged.
The inclusion of charge and long-ranged interactions is an active area of development~\cite{zhong_machine_2025,falletta_unified_2025,fang_phonon_2024}.
These developments are essential to addressing supercell scaling thereby enabling the study of the full range of possible defects in large supercells with our approach.

The T center is a best case scenario for a preliminary demonstration of the supercell scaling of our approach.
Si is a non-polar host material, and the ground state of the T center is electrically neutral.
The weak electron-phonon coupling of the T center and high purity of the Si host means that many intricate features are observed in the luminescence spectrum.
A discussion on supercell convergence needs to be framed within the context of required precision.
Defects with strong electron-phonon coupling produce broad luminescence;
for such cases, a modest supercell is sufficient.
This is why good agreement with experiment was obtained for several cases studied here, despite not addressing extrapolation to large supercells.

Finally, while the HSE hybrid functional was used throughout this work, we emphasize that this approach could be applied to any level of theory.
The significant computational savings of our approach makes even the random phase approximation~\cite{kaltak_cubic_2014}, time-dependent DFT~\cite{jin_excited_2023}, or wave-function methods~\cite{muechler_quantum_2022} tractable for predicting luminescence spectra.

In summary, we have demonstrated the fast, predictive power of MLIPs for the investigation of defect vibrational properties, particularly for luminescence spectra.
Utilizing the MACE framework, we established the ability of foundation models to produce qualitatively correct spectra and outlined a systematic approach for fine-tuning.
Fine-tuning to the atomic relaxation dataset enables predictions with accuracy rivaling that of the underlying DFT calculations.
When further precision is desired, additional configurations can be generated.
We demonstrated the efficiency and predictive capabilities of this approach by studying four defects in materials of diverse composition, culminating in an 8000-atom evaluation of the luminescence spectrum of the T center in Si, a prominent defect at the center of quantum computing efforts.
This work constitutes a significant step forward in the application of MLIPs to defects, and we are optimistic that efforts to improve MLIPs will enhance the ability to analyze and characterize defects from first principles.

\section*{Methods}
\subsection*{DFT Calculations}
We perform density functional theory calculations with version 6.4.3 of the VASP code~\cite{kresse_efficient_1996,kresse_efficiency_1996}.
For all of our DFT calculations, we utilize the hybrid functional of Heyd, Scuseria, and Ernzerhof~\cite{heyd_hybrid_2003,heyd_erratum:_2006}.
As noted in the introduction, the use of a hybrid functional is necessary to realize quantitative predictions for defects in solids.
Further details on the DFT benchmarking calculations are given in Supplementary Information \cref{sec:comp_det}.

\subsection*{Luminescence Theory}
\label{sec:theory}
We evaluate luminescence spectra using the first-principles formulation of Alkauskas \textit{et al.}~\cite{alkauskas_first-principles_2014-1}, which has become the standard technique for such calculations.
The approach uses the generating-function technique based on the works of Lax~\cite{lax_franckcondon_1952} and Kubo and Toyozawa~\cite{kubo_method_1954}.
Temperature dependence is included following the suggestions of Jin \textit{et al}~\cite{jin_photoluminescence_2021}.
Once an optical transition is identified, the mass-weighted atomic displacement vector $\Delta {\bm Q}$ between the ground and excited states is determined:
\begin{equation}
    \label{eq:dq}
    {\Delta Q}_{I\alpha} = \sqrt{M_I} \left( {R}^0_{e;I\alpha} - {R}^0_{g;I\alpha} \right) \;,
\end{equation}
where $I$ indexes the $N$ atoms and $\alpha$ the three Cartesian directions, $M_I$ is the atomic mass of the $I$th site, and ${\bm R}_{g/e}$ are the atomic coordinates of the ground ($g$) and excited ($e$) states with a superscript zero denoting the equilibrium geometry.
The atomic displacement vector is the main quantity in determining the coupling to phonons.

In addition to $\Delta {\bm Q}$, the phonon modes of the final state of the transition are required.
For luminescence, the final state of the transition corresponds to the ground state, whereas for absorption, the final state is the excited state.
We define the dynamical matrix ${\bm \Phi}$ as
\begin{equation}
    \label{eq:dynmat}
    \Phi_{I\alpha,J\beta} = \frac{1}{\sqrt{M_I M_J}} \frac{\partial^2 E_g}{\partial R_{g;I\alpha} \, \partial R_{g;J\beta}} \bigg\vert_{{\bm R}_g^0} \;,
\end{equation}
where $E_g$ is the total energy of the ground state.
The dynamical matrix is then diagonalized to obtain the phonon eigenvectors and eigenvalues,
\begin{equation}
    \label{eq:diag}
    {\bm \Phi} {\bm U} = {\bm U} {\bm \Omega}^2 \;,
\end{equation}
where ${\bm U} = ({\bm \eta}_1 \, \dots \, {\bm \eta}_{3N})$ is the orthogonal transformation matrix with the eigenvectors ${\bm \eta}_i$ as columns.
${\bm \Omega}^2 = {\rm diag}(\omega_1^2, \, \dots, \, \omega_{3N}^2)$ is the diagonal matrix of eigenvalues, which correspond to the phonon frequencies $\omega_i$ squared.

Each phonon mode contributes an amount $\Delta q_i$ to the atomic displacement, where
\begin{equation}
    \label{eq:dqi}
    \Delta q_i = {\bm \eta}_i \cdot \Delta {\bm Q}
\end{equation}
or in matrix form, $\Delta {\bm q} = {\bm U}^\mathsf{T} \, \Delta {\bm Q}$.
For each mode, we define a partial Huang-Rhys factor given by
\begin{equation}
    \label{eq:Si}
    S_i = \frac{1}{2\hbar} \omega_i {(\Delta q_i)}^2 \;,
\end{equation}
which can be interpreted as the average number of phonons with frequency $\omega_i$ emitted during the transition.

The total Huang-Rhys factor $S_{\rm tot}$ is of particular importance.
It is defined as
\begin{equation}
    \label{eq:Stot}
    S_{\rm tot} = \sum_i S_i \;.
\end{equation}
$S_{\rm tot}$ quantifies the total strength of electron-phonon coupling and is used to define the Debye-Waller factor $e^{-S_{\rm tot}}$, which describes the fraction of light emitted into the zero-phonon line (as opposed to the phonon sideband).

From the partial Huang-Rhys factors, we can define the spectral density of electron-phonon coupling $S(\hbar\omega)$, given by
\begin{equation}
    \label{eq:Shw}
    S(\hbar\omega) = \sum_i S_i \, \delta (\hbar\omega - \hbar\omega_i) \;,
\end{equation}
and its temperature-dependent counterpart
\begin{equation}
    \label{eq:Chw}
    C(\hbar\omega, T) = \sum_i \bar{n}_i (T) \, S_i \, \delta (\hbar\omega - \hbar\omega_i) \;,
\end{equation}
where $\bar{n}_i (T) = {\left[ \exp (\hbar\omega_i / k_{\rm B} T) - 1 \right]}^{-1}$ is the average occupation factor of the $i$th mode at temperature $T$.
In practice, the delta functions in \cref{eq:Shw,eq:Chw} are replaced with a Gaussian whose width varies linearly with the phonon frequency (Supplementary Information \cref{sec:broad}).
When local vibrational modes are present, a Lorentzian is used in place of the Gaussian function for each local vibrational mode, following the suggestion of Ref.~\onlinecite{silkinis_optical_2025}.
Local vibrational modes are identified by their inverse participation ratio (Supplementary Information \cref{sec:broad}).
Although it is not directly needed for calculating the luminescence spectrum, we define the phonon density of states as
\begin{equation}
    \label{eq:dos}
    \rho(\hbar\omega) = \frac{1}{3N} \sum_i \delta (\hbar\omega - \hbar\omega_i) \;,
\end{equation}
where the same broadening procedure described above is used to address the delta functions.
The phonon density of states will be used for benchmarking purposes.

Finally, the normalized luminescence intensity is obtained from
\begin{equation}
    \label{eq:Lhw}
    L(\hbar\omega, T) = C \, \omega^3 A(\hbar\omega, T) \;,
\end{equation}
where $C^{-1} = \int \omega^3 A(\hbar\omega, T) \, d(\hbar\omega)$ is a normalization constant.
$A(\hbar\omega, T)$ is an intermediate spectral function, obtained as:
\begin{equation}
    \label{eq:Ahw}
    A(\hbar\omega, T) = \frac{1}{2\pi} \int_{-\infty}^{\infty} e^{i\omega t - \gamma \lvert t \rvert - \sigma^2 t^2 / 2} \, G(t, T) \, dt \;,
\end{equation}
where $\gamma$ is included to account for homogeneous broadening and $\sigma$ for inhomogeneous broadening.
The generating function $G(t, T)$ is given by
\begin{multline}
    \label{eq:Gt}
    G(t, T) = \exp [ -i E_{\rm ZPL} t / \hbar + S(t) - S(0) \\
    + C(t, T) + C(-t, T) - 2 C(0, T) ] \;,
\end{multline}
where $S(t) = \int e^{i \omega t} S(\hbar\omega) \, d(\hbar\omega)$ and equivalently for $C(t, T)$. 
$E_{\rm ZPL}$ is the energy of the zero-phonon line.

The main computational bottleneck in evaluating the luminescence intensity $L(\hbar\omega, T)$ [\cref{eq:Lhw}] is obtaining the dynamical matrix ${\bm \Phi}$ [\cref{eq:dynmat}].
Typically ${\bm \Phi}$ is obtained using a finite-difference formula expressed in terms of the atomic forces, which are provided by the underlying DFT code.
For a cell with $N$ atoms, $N_{\rm FD} \times 3N$ fixed-geometry DFT calculations must be performed, where $N_{\rm FD}$ is the number of finite difference steps (typically two).
To resolve the intricate details of the phonon sideband, supercells containing of order 500 atoms are typically required~\cite{jin_photoluminescence_2021}.
Symmetry can reduce the number of required calculations in some cases, but in general, defects lower the symmetry of the system with respect to bulk, leaving thousands of calculations to be performed.
This prohibitive expense has driven researchers to utilize phonons calculated at a lower level of theory, which has been successful but comes with important caveats.
Typically, the atomic displacement vector $\Delta {\bm Q}$ still needs to be computed with a higher level of theory~\cite{alkauskas_first-principles_2014-1,jin_photoluminescence_2021}.
In addition, this approach only works for defects where the semi-local functional gets the description of the defect states qualitatively correct, which is not always the case~\cite{wickramaratne_assessing_2024}.

Motivated by these observation, we replace the evaluation of ${\bm \Phi}$ with the dynamical matrix obtained from a MLIP $\hat{\bm \Phi}$.
(A ``hat'' will be used to distinguish MLIP-predicted quantities when needed.)
We note that many implementations of MLIPs provide analytical evaluation of $\hat{\bm \Phi}$, avoiding the need for a finite-difference approximation.

\subsection*{Machine-Learning Interatomic Potentials}
For our MLIP framework, we employ the atomic cluster expansion with message passing (MACE)~\cite{batatia_mace_2022,batatia_design_2025}.
As the starting point, we utilize the \texttt{mace-omat-0-medium} foundation model~\cite{batatia_foundation_2024}, which is pre-trained on the Open Materials 2024 dataset~\cite{barroso-luque_open_2024}.
This dataset contains over 110 million DFT calculations on diverse structures with the PBE(+U) semi-local functional~\cite{perdew_generalized_1996,anisimov_band_1991}.
Although not explored here, we anticipate that our results are equally applicable to other MLIP frameworks.

We utilize the 64-bit floating point implementation of MACE.
As our goal is to obtain accurate dynamical matrix predictions, we set the force contribution to the loss to be 10 times that of the energy contribution.
The stress contribution is set to zero as the stresses are not evaluated on our dataset.
The \texttt{Adam} optimizer is used for training with a learning rate of 0.001 and a weight decay of 0.999.
We found the training to be fairly insensitive to these values, so long as the learning rate (weight decay) was not abnormally large (small).
Training is run for a maximum of 500 epochs.
The batch size has a pronounced effect on the training performance, as discussed in Supplementary Information \cref{sec:batch}, and we utilize a batch size of one here.
Isolated atom energies (``E0s'') are estimated by minimizing the prediction error assuming linearity~\cite{e0s};
we tested that this procedure does not influence model performance compared to using the explicitly calculated values.

During training, we do not utilize a train-test-validation split, given the limited number of configurations in our training dataset.
In principle, the DFT benchmarking calculations of the spectra (black lines in \cref{fig:bench_pl}) could be viewed as a test dataset.
We do not intend for equivalent benchmarking calculations to be done in practice when applying our approach, as that would defeat the purpose.

We also do not utilize ``multihead'' fine-tuning~\cite{mh}:
while multihead fine-tuning can help to stabilize the model, we found that the stabilization comes at the expense of better force prediction.
Although overfitting is always a concern, we do not see any evidence of it in our models for our intended application.
For this application, we are not concerned with the generalizability of the model, and evaluating the dynamical matrix requires a rather minimal exploration of the potential energy surface.
Indeed, the training choices made here may not be suitable for more sensitive applications like structure searching, molecular dynamics runs, or higher-order force constant evaluations.

Extensive analysis and benchmarking is performed for the various MLIP models utilized in this work, as discussed in Supplementary Information \cref{sec:benchmarking}.
Prior to computing the dynamical matrix, we perform an atomic relaxation with the model;
however, we note that this step was only necessary in the case of the foundation model.

\subsection*{Atomic Relaxation Dataset}
In performing the initial study of a defect with DFT, an atomic relaxation is performed.
This relaxation provides a small dataset of configurations (typically, 10--100) with energies and forces evaluated.
From the perspective of computing the luminescence with a MLIP, this data is effectively free, requiring no additional DFT calculations to be performed than would have already been necessary.
It is therefore a natural starting point for fine-tuning.
In addition to the relaxation data, a pristine supercell calculation is usually required;
as the supercell is constructed from a relaxed primitive cell, only a single configuration is evaluated.
Since this configuration is similarly ``free'', we include it in the training dataset.
Our relaxation datasets contain 48 configurations for C$_{\rm N}$ in GaN, 60 for the carbon dimer in hBN, and 27 for the NV center in diamond.
For the application examples, the relaxation datasets contain 19 configurations for N$_{\rm O}$ in ZnO, 23 for the divacancy in SiC, 10 for Bi$_{\rm Pb}$ in CsPbBr$_3$, and 41 for the T center in Si.
The maximum force on any atom in the dataset is 1.26~eV/{\AA} for the C$_{\rm N}$ in GaN dataset, 14.2~eV/{\AA} for the carbon dimer in hBN, 15.3~eV/{\AA} for the NV center in diamond, 1.86~eV/{\AA} for N$_{\rm O}$ in ZnO, 2.00~eV/{\AA} for the divacancy in SiC, 0.61~eV/{\AA} for Bi$_{\rm Pb}$ in CsPbBr$_3$, and 1.55~eV/{\AA} for the T center in Si.

We assess different methods for cleaning the relaxation dataset in Supplementary Information \cref{sec:rd_bench}.
Here we remove configurations where the maximum force on a given atom is above 2~eV/{\AA}.

\subsection*{Additional Data Generation}
Our goal is to generate additional configurations that provide an improved description of the phonons across all possible frequencies.
To accomplish this, we compute displacements within the phonon basis, where the phonons are obtained by diagonalizing the dynamical matrix predicted by the foundation model [\cref{eq:diag}].
While the foundation model does not exactly describe the phonons, it provides a qualitatively correct picture, differentiating between acoustic, optical, and local vibrational modes.

We refer to the following method as the ``optimized phonon'' method.
The phonon modes predicted by the foundation model are sorted into $N_{\rm conf}$ bins $\mathcal{B}_j$, grouping by frequency, where $N_{\rm conf}$ is the number of additional configurations desired.
A vector ${\bm \xi}_j$ is constructed in each bin as
\begin{equation}
    \label{eq:rand_phon}
    {\bm \xi}_j = \sum_{i \in \mathcal{B}_j} c_{ij} \hat{\bm \eta}_i \;,
\end{equation}
where the coefficients $c_{ij}$ are subject to normalization ($\sum_i \lvert c_{ij} \rvert^2 = 1$), and $\hat{\bm \eta}_i$ are the phonon eigenvectors of the foundation model dynamical matrix [\cref{eq:diag}].

For each random vector ${\bm \xi}_j$, the coefficients $c_{ij}$ are optimized to minimize the spread in forces on each atom, starting from a random distribution.
Specifically, we define a loss function $\mathcal{L}$ given by
\begin{equation}
    \label{eq:opt_phon}
    \mathcal{L} = \langle (\mathcal{\bm F}_j - \langle \mathcal{\bm F}_j \rangle)^2 \rangle \;,
\end{equation}
where the angled brackets denote an average over atoms, and $\mathcal{\bm F}_j$ is a normalized force magnitude due to random vector ${\bm \xi}_j$:
\begin{equation}
    \label{eq:fx}
    \mathcal{F}_{j;I}^2 = M_I \sum_{\alpha = x,y,z} \sum_i (c_{ij} \hat{\omega}_i^2 \hat{\eta}_{i;I\alpha})^2 \;.
\end{equation}
$\mathcal{L}$ is then minimized with respect to $c_{ij}$.
As each random vector ${\bm \xi}_j$ contributes independently to $\mathcal{L}$, they can be optimized individually.
The goal of this optimization procedure is to produce configurations with a similar force magnitude on each atom, which makes each atom in each configuration have a similar contribution to the training of the MLIP.

We assess alternative methods of generating additional configurations and benchmark them in Supplementary Information \cref{sec:ad_bench}.
All of the explored methods exhibited similar performance, but we find a small advantage to using the optimized phonon method.
We also explore using configurations from a pristine bulk supercell in Supplementary Information \cref{sec:ad_bench}.

\subsection*{Large Supercell T Center Calculations}
To obtain the precision necessary to fully compare with experiments on the T center, we need to address supercell scaling.
We first evaluate the atomic displacement vector $\Delta {\bm Q}$ in a ``small'' 512-atom supercell.
As the T center is a bound-exciton emitter, we utilize the negative charge state to approximate the geometry of the bound-exciton excited state, following the suggestion of Ref.~\onlinecite{silkinis_optical_2025}.
We fine-tune a MACE model to the relaxation dataset of the ground state, 30 additional configurations generated using the optimized phonon method detailed above, and 10 random displacements in the pristine supercell.

The MACE dynamical matrix is then used to construct the forces that would give rise to this displacement within the harmonic approximation~\cite{turiansky_approximate_2025}.
These forces are then zero-padded to embed them into a large 8000-atom supercell.
This takes advantage of the fact that these forces are more short-ranged than the displacement vector itself~\cite{alkauskas_first-principles_2014-1,razinkovas_vibrational_2021}.
We then use our fine-tuned MACE model to evaluate the dynamical matrix in the 8000-atom supercell, without any further fine-tuning or DFT calculations.
The conventional embedding scheme that exists in the literature constructs the dynamical matrix by patching together the dynamical matrices of ``small'' defective and pristine supercells~\cite{alkauskas_first-principles_2014-1,razinkovas_vibrational_2021}.
By utilizing MLIPs, we are able to avoid this approximation.

The luminescence spectra calculated in a 512-atom cell without embedding and in the 8000-atom cell with embedding are compared in Supplementary Information \cref{sec:small_cell}.

\section*{Data Availability}
The data that supports the findings of this study are available from the corresponding author upon reasonable request.

\section*{Code Availability}
The code implementing the methodology used in this study is available via Github at \url{https://github.com/mturiansky/lineshape_tools} (Ref.~\onlinecite{code}).

\begin{acknowledgments}
    We acknowledge fruitful discussions with I.~Batatia and J.~Hart.
    This work was supported by the Office of Naval Research through the Naval Research Laboratory's Basic Research Program.
    Calculations were supported by high-performance computer time and resources from the DoD High Performance Computing Modernization Program.
\end{acknowledgments}

\end{document}


\title{Supplementary Information for\\Machine Learning Phonon Spectra for\\Fast and Accurate Optical Lineshapes of Defects}

\author{Mark E. Turiansky}
\email{mark.e.turiansky.ctr@us.navy.mil}
\affiliation{US Naval Research Laboratory, 4555 Overlook Avenue SW, Washington, DC 20375, USA}

\author{John L. Lyons}
\affiliation{US Naval Research Laboratory, 4555 Overlook Avenue SW, Washington, DC 20375, USA}

\author{Noam Bernstein}
\affiliation{US Naval Research Laboratory, 4555 Overlook Avenue SW, Washington, DC 20375, USA}

\date{\today}

\maketitle

\section{Computational Details}
\label{sec:comp_det}
We perform density functional theory calculations with version 6.4.3 of the VASP code~\cite{kresse_efficient_1996,kresse_efficiency_1996}.
Core electrons are frozen within the projector augmented-wave formalism~\cite{blochl_projector_1994}, and the valence electrons are represented in a plane-wave basis.
The basis is truncated at an energy of 400~eV for GaN, ZnO, and Si, 500~eV for CsPbBr$_3$, and 520~eV for h-BN, diamond, and SiC.
Ga and Pb $d$ states are included in the core, while Bi $d$ and Cs $s$ states are included as valence.
We utilize the hybrid functional of Heyd, Scuseria, and Ernzerhof~\cite{heyd_hybrid_2003,heyd_erratum:_2006} to ensure accurate electronic structure.
The fraction of short-range Hartree-Fock exchange is set to 0.31 for GaN, 0.35 for CsPbBr$_3$, 0.36 for ZnO, 0.40 for h-BN, and 0.25 for diamond, Si, and SiC.
The default screening parameter of 0.2~{\AA}$^{-1}$ is utilized for all materials except CsPbBr$_3$, for which we use 0.1~{\AA}$^{-1}$.
For h-BN, we empirically include van der Waals interactions via the Grimme-D3 scheme~\cite{grimme_consistent_2010} with the default parameters for PBE, which were also the default for HSE prior to VASP version 6.
These values are consistent with previous studies~\cite{turiansky_dangling_2019,mackoit-sinkeviciene_carbon_2019,lyons_first-principles_2017,alkauskas_first-principles_2014-1,gali_ab_2019,lyons_carbon_2010,lyons_effects_2014} and chosen to reproduce the experimental band gaps.

Defects are studied in a supercell configuration within periodic boundary conditions~\cite{freysoldt_first-principles_2014}.
We utilize a 96-atom supercell for GaN and ZnO, which is a $3\times2\times2$ multiple of the 8-atom orthorhombic cell obtained by transforming the 4-atom primitive cell.
For h-BN, we similarly transform the primitive cell into an 8-atom orthorhombic cell and scale to a $5\times3\times2$ multiple, which contains 240 atoms.
A 216-atom supercell of diamond is constructed as a $3\times3\times3$ multiple of the conventional cubic unit cell.
For Si, we utilize a 512-atom supercell, which is a $4\times4\times4$ multiple of the conventional cubic unit cell.
We generate a 512-atom supercell of SiC using the optimized supercell generation routine in the \texttt{doped} code~\cite{kavanagh_doped_2024}.
Lastly, we study an 80-atom supercell of CsPbBr$_3$.
For details on the atomic geometries and electronic structure of the defects, we refer the reader to Refs.~\onlinecite{lyons_carbon_2010,lyons_effects_2014,alkauskas_first-principles_2014-1,gali_ab_2019,mackoit-sinkeviciene_carbon_2019,turiansky_approximate_2025}.

The Brillouin zone is sampled with the mean-value point~\cite{baldereschi_mean-value_1973} for all calculations, with the exception of Si and SiC for which the supercells are large enough to use the $\Gamma$ point.
Atomic coordinates are relaxed until the forces are below 5~meV/{\AA}.
The lattice parameters are kept fixed at the calculated bulk values.
Spin polarization is explicitly taken into account.

The transitions investigated for the NV center in diamond, the carbon dimer in hBN, and the divacancy in SiC correspond to the lowest energy spin-conserving internal transitions.
To address the excited states of an internal transition, we utilize the constrained-occupation $\Delta$SCF approach~\cite{jones_density_1989}.
In the case of the carbon dimer, the excited state is multideterminant, and we correct the ZPL energy as was done in Ref.~\onlinecite{mackoit-sinkeviciene_carbon_2019}, following the suggestion of von Barth~\cite{von_barth_local-density_1979}.
However, we approximate the potential energy surface of the excited state with the potential energy surface of the mixed state, consistent with Ref.~\onlinecite{mackoit-sinkeviciene_carbon_2019}.
For the NV center in diamond and the divacancy in SiC, we utilize the low-symmetry electronic occupation for the excited state, which results from a Jahn-Teller distortion.
A complete study of the luminescence properties should explicitly account for the Jahn-Teller effect~\cite{bersuker_jahn-teller_2006,razinkovas_vibrational_2021,zalandauskas_theory_2025}, but this is beyond the scope of the present study.

\section{Broadening}
\label{sec:broad}
As discussed in the main text, the delta functions in the spectral density $S(\hbar\omega)$ [\cref{eq:Shw}] are replaced with a Gaussian, or in the case of local vibrational modes, a Lorentzian.
We vary the broadening parameter of the Gaussian linearly from $\sigma_{\rm low}$ at zero to $\sigma_{\rm high}$ at the highest phonon frequency (only considering phonon modes that are not local vibrational modes).
Local vibrational modes are identified by their inverse participation ratio, defined as
\begin{equation}
    \label{eq:ipr}
    \frac{1}{{\rm IPR}_i} = \sum_I \left(\sum_{\alpha=x,y,z} \eta_{i;I\alpha}^2\right)^2 \;,
\end{equation}
where ${\bm \eta}_i$ are the normalized phonon eigenvectors [\cref{eq:diag}].
For the T center, modes with ${\rm IPR}_i < 800$ are broadened by a Lorentzian with a width of 0.15~meV (instead of a Gaussian).
In the evaluation of the luminescence, $\gamma$ accounts for homogeneous broadening and $\sigma$ for inhomogeneous broadening [see \cref{eq:Ahw}].
The values that were used to generate the luminescence spectra in the main text are specified in \cref{tab:broad}.

\begin{table}[ht!]
    \centering
    \caption{\label{tab:broad}
        The low $\sigma_{\rm low}$ and high $\sigma_{\rm high}$ Gaussian broadening parameters in the spectral density, as well as the homogeneous $\gamma$ and inhomogeneous $\sigma$ broadening parameters in the spectral function.
    }
    \begin{ruledtabular}
       \begin{tabularx}{\columnwidth}{c S[table-format=1.2]S[table-format=1.2]S[table-format=1.2]S[table-format=1.2]} 
            Defect&{$\sigma_{\rm low}$ [meV]}&{$\sigma_{\rm high}$ [meV]}&{$\gamma$ [meV]}&{$\sigma$ [meV]} \\
            \midrule
            C$_{\rm N}$ in GaN&3.5&1.0&2.5&0 \\
            NV in diamond&7.5&2.5&3.0&0 \\
            C$_{\rm B}$-C$_{\rm N}$ in h-BN&20&7.0&7.5&0 \\
            N$_{\rm O}$ in ZnO&5.0&1.0&2.0&55 \\
            $V_{\rm Si}V_{\rm C}$ in SiC&4.0&1.5&0.1&0 \\
            Bi$_{\rm Pb}$ in CsPbBr$_3$&1.2&1.2&2.0&55 \\
            T Center in Si&1.5&1.0&0.15&0 \\
        \end{tabularx}
    \end{ruledtabular}
\end{table}

\section{Batch Size Dependence}
\label{sec:batch}
\begin{figure}[htb!]
    \centering
    \includegraphics[width=0.5\columnwidth,height=0.5\textheight,keepaspectratio]{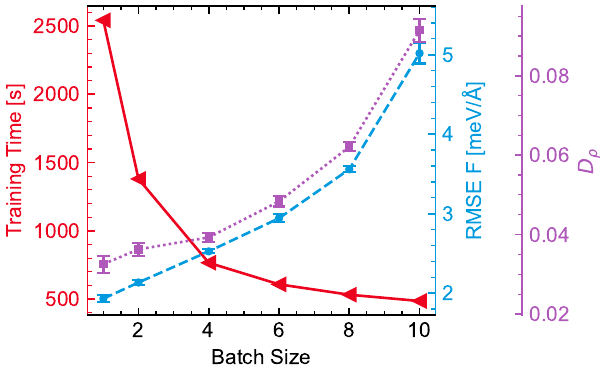}
    \caption{\label{fig:batch_size}
        Dependence of training time (red, left triangle, solid), root mean square error in the forces RMSE$(F)$ (blue, circle, dashed), and density of states error $D_\rho$ (purple, square, dotted) as a function of batch size during training for C$_{\rm N}$ in GaN.
        Error bars correspond to the standard deviation of ten trainings with different seeds.
        Training times are from a single, representative training.
    }
\end{figure}

The parameter that we found to have the largest effect on the MLIP training was the batch size, which controls the number of samples used in updating the gradient during training.
As shown in \cref{fig:batch_size} for C$_{\rm N}$ in GaN, decreasing the batch size improved the model performance, exemplified by RMSE$(F)$ and $D_\rho$ (defined in \cref{sec:crit}).
This comes at the cost of increased training time.
Ultimately, training to our relatively small dataset took no longer than an hour on a single NVIDIA A100 GPU in the cases we explored.
For the present work, we utilize a batch size of one, as it performs the best.
It may be useful to increase the batch size to four in big-data applications where some accuracy can be sacrificed for speed.

\section{MLIP Benchmarking}
\label{sec:benchmarking}

\subsection{Benchmarking Criteria}
\label{sec:crit}
To quantify the performance of the MLIP at reproducing the explicit DFT calculations, we will focus on five metrics computed assuming $T = 0$.
First is the maximum force magnitude on a given atom predicted by the MLIP in the DFT equilibrium geometry $F_0 = \max_I \, \{ \lvert \hat{\bm F}_I^0 \rvert \}$, where $\hat{F}_{I\alpha}^0 = \partial \hat{E}_g (\{{\bf R}_g^0\}) / \partial R_{g;I\alpha}$.
This quantifies the degree to which the MLIP has learned the equilibrium geometry.
Second is the root mean square error of the force prediction:
\begin{equation}
    \label{eq:rmsef}
    {\rm RMSE}(F) = \frac{1}{N_c} \sum_c \frac{1}{3 N} \sum_{I\alpha} (\hat{F}_{c;I\alpha} - F_{c;I\alpha})^2 \;,
\end{equation}
where $c$ is summed over the $N_c$ configurations in the dataset.
$F_{c;I\alpha}$ is the force on atom $I$ along Cartesian direction $\alpha$ in atomic configuration $c$.
Third is the relative error in the total Huang-Rhys factor ${\rm RE}({S_{\rm tot}}) = \lvert \hat{S}_{\rm tot} - S_{\rm tot} \rvert / S_{\rm tot}$.

The remaining metrics quantify how well the shapes of the phonon density of states $\rho (\hbar\omega)$ and the electron-phonon spectral density $S(\hbar\omega)$ are described.
For a quantitative measure of this error, we use the Bhattacharyya distance~\cite{bhattacharyya_measure_1946}, given by
\begin{equation}
    \label{eq:D}
    D_f = -\ln \left( \int \sqrt{\hat{f} (\hbar\omega) \; f(\hbar\omega)} \, d(\hbar\omega) \right) \;,
\end{equation}
where $f(\hbar\omega)$ is the normalized distribution obtained from $\rho(\hbar\omega)$ or $S(\hbar\omega)$.
A smaller value of $D_f$ indicates better agreement between the prediction and true value.
Each order of magnitude that $D_f$ decreases corresponds to an additional ``9'' in the overlap integral [i.e., $-\!\ln(0.99) \approx 0.01$, while $-\!\ln(0.9999) \approx 0.0001$].

The phonon density of states $\rho(\hbar\omega)$ is explicitly related to the phonon frequencies, and $D_{\rho}$ therefore quantifies the accuracy with which the phonon frequencies are predicted.
On the other hand, the electron-phonon spectral density $S(\hbar\omega)$ encodes information not only on the phonon frequencies, but also on the eigenvectors, through the projection of the atomic displacement vector into the phonon basis [$\Delta q_i$, \cref{eq:dqi}].
$D_S$ therefore quantifies the quality with which the phonon eigenvectors are predicted.
In passing, we note that it would be useful to explore defining an equivalent $D_S$ for characterizing the eigenvectors of pristine, bulk phonons.
By selecting a transition at a defect, we are given a unique vector that can be projected into the phonon basis.
A random or unit displacement vector may be a suitable alternative for assessing bulk phonons.

These five quantities ($F_0$, RMSE$(F)$, RE$(S_{\rm tot})$, $D_\rho$, and $D_S$) capture the essential information necessary to predict the vibrational properties of a given defect and therefore the luminescence.
One may wonder why we do not also define $D_L$ for the luminescence spectrum, which is ultimately the main quantity that we are trying to predict.
We found $D_L$ to be almost completely determined by RE$(S_{\rm tot})$ and $D_S$, and therefore, specifying it provides no additional information.

As some aspects of training are stochastic, we performed 10 trainings with different seeds for the random number generator for benchmarking purposes unless otherwise noted.
When error bars are shown, we report the mean and standard deviation of our benchmark quantities for these trainings.
Otherwise the default seed value of MACE (i.e., 123) is used.

\subsection{Foundation Model}
\label{sec:fm_bench}
We first assess the performance of the baseline foundation model for predicting luminescence spectra.
With the foundation model, we compute $F_0$ and find a value of 1.52 eV/{\AA} for C$_{\rm N}$ in GaN, 1.00 eV/{\AA} for the NV center in diamond, and 0.97 eV/{\AA} for the carbon dimer in hBN (see \cref{tab:bench}).
Unsurprisingly, the foundation model does not predict the equilibrium geometry at a force convergence criteria of 5 meV/{\AA}, which is the value used for the DFT relaxations.
Prior to computing the dynamical matrix, we perform an atomic relaxation with the foundation model.
We note that this atomic relaxation with the foundation model does not influence the determination of the relaxation vector $\Delta {\bm Q}$ used in evaluating the optical spectra;
its purpose is to obtain a reliable dynamical matrix that approximates the actual dynamical matrix computed in the true equilibrium geometry.

\begin{table*}[ht!]
    \centering
    \caption{\label{tab:bench}
        The model forces evaluated on the DFT equilibrium geometry $F_0$, root mean square error in forces RMSE$(F)$, density of states error $D_\rho$, spectral density error $D_S$, and relative error in the Huang-Rhys factor RE$(S_{\rm tot})$.
        Foundation model describes the out-of-the-box performance of the \texttt{mace-omat-0-medium} foundation model.
        Relax data corresponds to fine-tuning to the atomic relaxation dataset only.
        Optim. Phon. corresponds to fine-tuning to the atomic relaxation dataset supplemented by additional configurations generated by the optimized phonon method with the number of additional configurations given in parenthesis.
        Lastly, Pristine corresponds to fine-tuning to the atomic relaxation dataset supplemented by ten additional configurations generated by randomly displacing a pristine supercell.
    }
    \begin{ruledtabular}
        \begin{tabularx}{\textwidth}{c c S[table-format=4.1]S[table-format=1.2]S[table-format=1.5]S[table-format=1.4]S[table-format=1.2]}
            Defect&Method&{$F_0$ [meV/{\AA}]}&{RMSE$(F)$ [meV/{\AA}]}&{$D_\rho$}&{$D_S$}&{RE$(S_{\rm tot})$ [\%]} \\
            \midrule
            \multirow{6}{*}{C$_{\rm N}$ in GaN}&Foundation Model&1520&175&0.14&0.067&3.82 \\
            &Relax Data&4.7&2.98&0.036&0.044&2.96 \\
            &Optim. Phon. (10)&4.2&1.57&0.00057&0.014&1.78 \\
            &Optim. Phon. (30)&3.6&1.30&0.00028&0.0020&1.05 \\
            &Optim. Phon. (100)&2.7&0.98&0.00018&0.0014&0.47 \\
            &Pristine (10)&4.4&1.96&0.0024&0.034&1.75 \\
            \midrule
            \multirow{5}{*}{NV in diamond}&Foundation Model&1000&127&0.080&0.044&12.4 \\
            &Relax Data&4.4&0.93&0.0014&0.0024&1.94 \\
            &Optim. Phon. (10)&5.0&0.57&0.000090&0.0015&1.53 \\
            &Optim. Phon. (30)&5.0&0.48&0.000079&0.00056&0.56 \\
            &Pristine (10)&4.4&0.69&0.00012&0.0023&2.17 \\
            \midrule
            \multirow{5}{*}{Carbon dimer in hBN}&Foundation Model&970&81.5&0.0064&0.023&0.49 \\
            &Relax Data&5.9&1.40&0.00023&0.0025&1.26 \\
            &Optim. Phon. (10)&7.0&1.20&0.000017&0.0022&1.35 \\
            &Optim. Phon. (30)&5.9&1.06&0.000055&0.0031&0.43 \\
        \end{tabularx}
    \end{ruledtabular}
\end{table*}

C$_{\rm N}$ in GaN is a case of strong electron-phonon coupling, characterized by an $S_{\rm tot}$ of 9.62.
Such a large value of $S_{\rm tot}$ means the luminescence intensity is broad and featureless, resembling a Gaussian function.
The foundation model predicts a qualitatively correct spectral density ($D_S = 0.067$, \cref{tab:bench}) with a main peak near $\approx$35~meV, while there are some deviations in the high frequency phonons consistent with the errors in the phonon density of states.
In addition, $\hat{S}_{\rm tot}$ takes a value of 9.25, within 3.82\% of the true value.
This underestimation of $S_{\rm tot}$ results in a peak luminescence intensity that is slightly shifted in energy from the true value.
When comparing with experiments, it is common to align the spectrum by applying a small energy shift to account for typical DFT errors.
In practice, the foundation model prediction is already sufficiently good for interpreting experiments.

The use of a hybrid functional to determine the displacement vector was essential for C$_{\rm N}$:
calculations based on semi-local functionals miss the localization of the hole wavefunction and corresponding asymmetric distortion of the atomic configuration~\cite{deak_optimized_2019,wickramaratne_assessing_2024}.
This missing physics led researchers to misinterpret C$_{\rm N}$ as a shallow dopant~\cite{lyons_carbon_2010}.
Moreover, the incorrect description of the asymmetric distortion resulting from localization precludes an accurate description of the optical properties of C$_{\rm N}$.

In the opposite extreme, the carbon dimer in hBN has relatively weak electron-phonon coupling ($S_{\rm tot} = 1.87$).
As a result, the luminescence is strongly peaked at the zero-phonon line energy.
The foundation model predictions for the carbon dimer show a rather remarkable agreement with the explicitly calculated ones.
Indeed, we find $D_S = 0.023$ and $\hat{S}_{\rm tot} = 1.87$ (see \cref{tab:bench}).
The luminescence is both qualitatively and quantitatively captured, with the phonon replicas described to within a few meV.
It has been previously noted that the electron-phonon coupling of the carbon dimer is somewhat simple and could be captured with a single-mode model~\cite{turiansky_approximate_2025}.
In addition, B, C, and N are first row elements, which tend to be well-represented in foundation model training sets.

Lastly, the NV center in diamond exhibits moderate electron-phonon coupling ($S_{\rm tot} = 3.47$) and some fine structure appears in the luminescence intensity.
While the qualitative shape of the spectral density is captured ($D_S = 0.044$, \cref{tab:bench}), the dominant peak is shifted to a lower frequency compared to the true calculation.
This affects the luminescence prediction, resulting in the phonon replicas in the phonon sideband being shifted closer to the zero-phonon line.
The foundation model predicts $\hat{S}_{\rm tot} = 3.04$, which is underestimated compared to the true value.
This result is less favorable in terms of enabling a comparison with experiments.
A rigid shift of the spectrum will not be able to fix the positions of the phonon replicas with respect to the zero-phonon line.
It would therefore be desirable to fine-tune the foundation model to enable a more meaningful comparison with experiments.

\subsection{Relaxation Data Fine-Tuning}
\label{sec:rd_bench}
In some cases, the relaxation dataset contains structures with large forces, which may be less relevant for evaluating phonons in the harmonic approximation.
In addition, the final steps of atomic relaxation involve evaluating several structures close to the equilibrium geometry, which may not be sufficiently distinct and contribute noise.
Towards this end, we assess three cleaning procedures.
One procedure is a simple force filter that removes configurations with forces larger than 2 eV/{\AA}.
A second is computing the displacement vector between configurations and discarding those whose atoms have not moved sufficiently far (a displacement norm less than 0.1~{\AA}).
The final procedure entails computing the normalized displacement vector with respect to the equilibrium geometry for each configuration.
A pivoted $QR$ factorization is then performed on these vectors to identify the $M$ most linearly independent vectors;
$M$ is the rank of the vector space determined by counting the diagonal elements of $R$ whose magnitude are within five orders of magnitude of the greatest value.

\begin{figure}[htb!]
    \centering
    \includegraphics[width=0.5\columnwidth,height=\textheight,keepaspectratio]{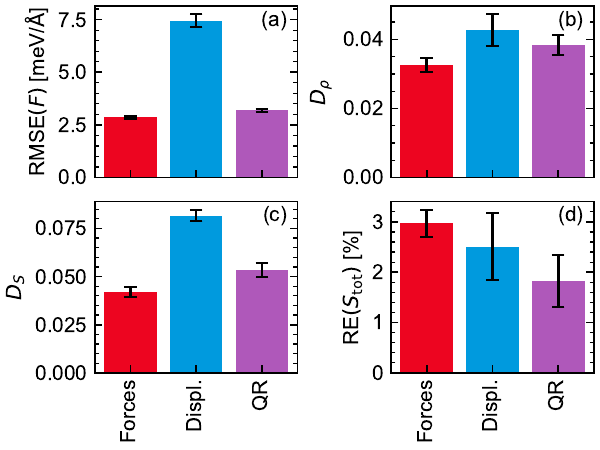}
    \caption{\label{fig:clean}
        The (a) root mean square error in forces RMSE$(F)$, (b) density of states error $D_\rho$, (c) spectral density error $D_S$, and (d) relative error in the Huang-Rhys factor RE$(S_{\rm tot})$.
        Error bars correspond to the standard deviation of ten trainings with different seeds.
        ``Forces'' (red) refer to filtering based on forces, ``Displ.'' (blue) refers to filtering based on the displacement vector, and ``QR'' (purple) refers to filtering based on a QR factorization to determine linear independence.
    }
\end{figure}

The results of the cleaning procedures are shown in \cref{fig:clean} for C$_{\rm N}$ in GaN.
Both the second and third procedures remove a large number of structures, and while this speeds up the training, it also degrades the performance.
Ultimately we conclude that, in this limited data regime, more data is better.
For this work and the following, we utilize the force filter.

In all benchmarking cases, the model fine-tuned to the relaxation dataset has effectively learned the DFT equilibrium geometry, as indicated by $F_0$ being close to or below 5~meV/{\AA} (\cref{tab:bench}).
As a consequence, the atomic relaxation performed prior to computing the dynamical matrix---which was necessary for the foundation model---is no longer needed.

The predicted luminescence spectra are significantly closer to the true DFT calculation for all three cases [\cref{fig:bench_pl}(c,f,i)].
For the carbon dimer and the NV center in diamond, the phonon density of states and spectral density are significantly improved.
When one looks closely at the spectra for the NV center in diamond, it is interesting that while the density of states is nearly perfectly captured in the vicinity of $\approx$60~meV, the spectral density still shows minor discrepancies.
This indicates that the phonon eigenvectors are not as well captured as the distribution of eigenvalues, pointing to remaining room for improvement.
The improvement for C$_{\rm N}$ in GaN is more modest;
in particular, there are remaining discrepancies in the high-frequency phonon modes, as seen in the phonon density of states [\cref{fig:bench_pl}(a)].
(These are ultimately washed out in the predicted luminescence due to the strong electron-phonon coupling, as noted above.)

\subsection{Generating Additional Data for Fine-Tuning}
\label{sec:ad_bench}
We explore three methods of generating additional data to further refine our model.
As the first method (labeled as ``Rand. Cart.''), we generate random displacement vectors and orthogonalize them with a Gram-Schmidt procedure.
Random vectors are conceptually simple but are not guaranteed to evenly sample the space of phonons:
the effective phonon frequency of a random phonon mode tends to be at higher frequencies, and low frequency portions of the spectrum will be missed.

To overcome this, we utilize the foundation model dynamical matrix to generate displacements within the phonon basis.
As shown in the main text and \cref{sec:fm_bench}, the foundation model does not provide an exact description of the phonon modes, but it does capture the major qualitative behaviors.
The second method (labeled as ``Rand. Phon.'') entails computing the phonon basis by diagonalizing the dynamical matrix predicted by the foundation model $\hat{\bm \Phi}$.
We then sort and divide the phonon modes into $N_{\rm conf}$ bins $\mathcal{B}_j$, grouping by frequency, where $N_{\rm conf}$ is the number of additional configurations desired.
A random vector ${\bm \xi}_j$ is then constructed in each bin as
\begin{equation}
    \label{eq:rand_phon_si}
    {\bm \xi}_j = \sum_{i \in \mathcal{B}_j} c_{ij} \hat{\bm \eta}_i \;,
\end{equation}
where $c_{ij}$ are random variables, subject to normalization ($\sum_i \lvert c_{ij} \rvert^2 = 1$), and $\hat{\bm \eta}_i$ are the phonon eigenvectors of the foundation model dynamical matrix [see \cref{eq:diag}].
The third method is the optimized phonon method (labeled as ``Optim. Phon.'') described in the main text (Methods).

We also tested a fourth method in which a single phonon mode $\hat{\bm \eta_i}$ was randomly selected from each bin $\mathcal{B}_j$.
As this method did not show any benefit over the other methods, we do not discuss it further.

\begin{figure*}[htb!]
    \centering
    \includegraphics[width=\textwidth,height=\textheight,keepaspectratio]{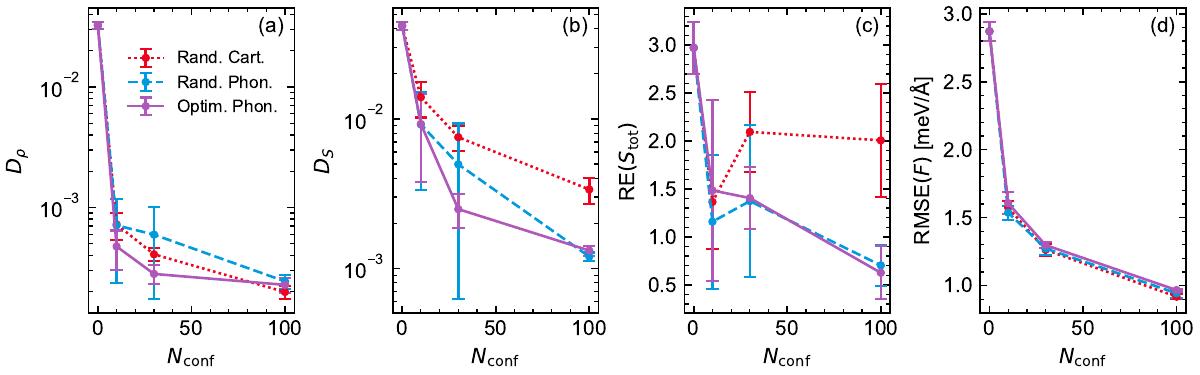}
    \caption{\label{fig:confs_error}
        (a) Density of state error $D_\rho$, (b) spectral density error $D_S$, (c) relative error of the Huang-Rhys factor RE$(S_{\rm tot})$, and (d) root mean square error in forces RMSE$(F)$ as a function of the additional configurations included in training $N_{\rm conf}$.
        The results that are shown are for C$_{\rm N}$ in GaN.
        The lines correspond to the three different methods of generating data described in the text Rand. Cart. (red, dotted), Rand. Phon. (blue, dashed), and Optim. Phon. (purple, solid).
        Error bars correspond to the standard deviation of 10 trainings for $N_{\rm conf} = 0$ and 100, 50 trainings for $N_{\rm conf} = 10$, 30 trainings for $N_{\rm conf} = 30$.
    }
\end{figure*}

As C$_{\rm N}$ in GaN proved most difficult to fine-tune in the preceding section, we use it to evaluate these three methods of generating additional data.
In \cref{fig:confs_error}, we show the $D_\rho$, $D_S$, $S_{\rm tot}$ and RMSE$(F)$ for $N_{\rm conf} = 0$, $10$, $30$, and $100$, where zero indicates training using only the relaxation dataset.
For these tests, $F_0$ was found to be below the force convergence threshold of 5~meV/{\AA}.
In addition to the inherent randomness of training, these three methods also include a random element in generating the displacement vectors.
For $N_{\rm conf} = 10$, we generate five datasets and do ten trainings each, resulting in 50 total trainings to average over.
Similarly for $N_{\rm conf} = 30$, we generate three datasets.

For all three methods, a significant improvement in the benchmarking criteria is obtained relative to using only the relaxation dataset.
Already with just ten additional configurations, $D_\rho$ has improved by more than an order of magnitude.
The improvement in $D_S$, which relies not only on phonon frequencies but also the eigenvectors, is more modest but still substantial.
Both quantities further improve with more configurations.

We find that there is a small, statistically significant benefit to utilizing the third, optimized phonon method ($p = 0.03$ for $D_\rho$ and 0.07 for $D_S$~\footnote{This $p$-value is computed assuming the number of samples is 5, the number of unique datasets, rather than 50, the number of trainings.} for $N_{\rm conf} = 10$), particularly for predicting the spectral density.
In practice, the difference between the three methods is rather negligible, as there is little qualitative difference between the predicted spectra.
The first method (random vectors) is also easier to converge with respect to the number of modes, as the methods that utilize phonon information require regenerating all the modes when $N_{\rm conf}$ is changed.
In the main text, we utilize the optimized phonon method (method three).

The improvement of the model from the inclusion of additional configurations raises the question of whether the initial atomic relaxation dataset is necessary.
We explicitly tested this possibility and found that, while the density of states was still well described, the prediction of the spectral density and resulting luminescence is degraded.
Indeed there is important information in the atomic relaxation data.

Another hypothesis that we tested was whether the additional configurations need to be from the defective supercell.
We tested this hypothesis by generating ten randomly displaced configurations in the pristine supercell that is the same size as the defective supercell.
When combined with the atomic relaxation dataset, the model performs almost as well as with the optimized phonon method (labeled ``Pristine'' in \cref{tab:bench}).
Generating configurations in the pristine supercell has two major advantages:
the pristine supercell is usually cheaper to calculate, due to its higher symmetry and the fact that spin polarization is not typically needed, and the configurations can be reused for training of other defects in the same material.
We would recommend doing an initial screening of defects in a material by fine-tuning to the atomic relaxation dataset combined with pristine supercell configurations.
When interesting cases are identified, additional configurations in the defective supercell can be generated.


\section{T Center Supercell Scaling}
\label{sec:small_cell}
\begin{figure}[htb!]
    \centering
    \includegraphics[width=0.5\columnwidth,height=\textheight,keepaspectratio]{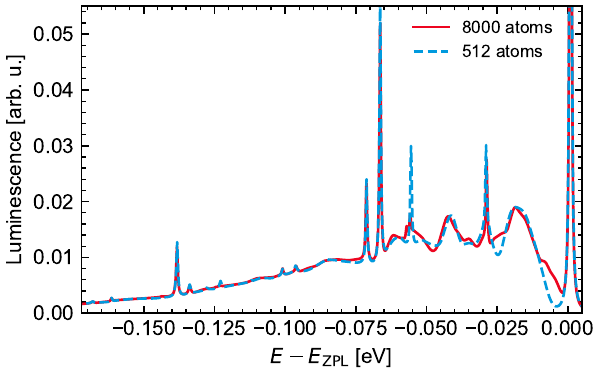}
    \caption{\label{fig:small_cell}
        Theoretical luminescence spectrum of the T center in Si at $T = 4.2$~K in a 512-atom (blue, dashed) and 8000-atom (red, solid) supercell.
    }
\end{figure}

The T center vibrational spectrum contains several local vibrational modes that overlap with the continuum of bulk modes.
These local vibrational modes, and the low-energy portion of the spectrum in general, show a more significant dependence on the supercell size.
In \cref{fig:small_cell}, we show the luminescence spectrum in a 512-atom supercell compared to the spectrum from the 8000-atom supercell shown in the main text (\cref{fig:Tcenter}).
For the 512-atom supercell, the Gaussian broadening parameters were $\sigma_{\rm low} = 2.5$~meV and $\sigma_{\rm high} = 2.0$~meV.
Local vibrational modes were identified by ${\rm IPR}_i < 100$, and all other broadening parameters were the same (see \cref{tab:broad}).
The intensity of the feature at 56~meV below the ZPL is significantly overestimated.
While the intensity is closer for the feature at $\approx$29~meV below the ZPL, the frequency is slightly shifted from the larger supercell.

\section{Proportionality of $S_{\rm tot}$ and the Mass of the Host Lattice}
\label{sec:stot_prop}
The strength of electron-phonon coupling as quantified by the Huang-Rhys factor $S_{\rm tot}$ [\cref{eq:Stot}] depends implicitly on the mass of the host lattice.
This can be shown by considering the Huang-Rhys factor in the 1D approximation, which is an upper bound on $S_{\rm tot}$~\cite{turiansky_approximate_2025}:
\begin{equation}
    \label{eq:S1d}
    S_{\rm 1D} = \frac{1}{2\hbar} \sqrt{2 W_{\rm tot}} \, \Delta Q \;,
\end{equation}
where $W_{\rm tot}$ is the relaxation energy and $\Delta Q = \Vert \Delta {\bm Q} \Vert$ [\cref{eq:dq}].
If we consider two defects with the same $W_{\rm tot}$ in different host materials, then $S_{\rm 1D} \propto \sqrt{M}$ [see \cref{eq:dq}], where $M$ is the reduced mass of the atomic motion along the displacement vector.
The atomic displacement vector typically involves motion of host atoms, and $S_{\rm 1D}$ will scale with the mass of the host lattice.
A heavier lattice therefore implies stronger electron-phonon coupling.
When the strength of electron-phonon coupling is larger, the luminescence is broad, and the fine details of the phonon sideband are washed out.
Bi$_{\rm Pb}$ in CsPbBr$_3$ was an exemplary case of this effect.